\documentclass[showkeys,showpacs,prd]{revtex4}
\usepackage{amsmath}
\usepackage{amssymb}
\usepackage{graphicx}
\usepackage{color}
\usepackage{ulem}

\begin{document}

\title{
Quantum torsion with non-zero standard deviation: non-perturbative approach for  cosmology
}

\author{
Vladimir Dzhunushaliev,$^{1,2,3,4}$
\footnote{Email: v.dzhunushaliev@gmail.com}
Vladimir Folomeev,$^{3,4}$
\footnote{Email: vfolomeev@mail.ru}
Burkhard Kleihaus,$^{4}$
\footnote{Email: b.kleihaus@uni-oldenburg.de }
Jutta Kunz$^4$
\footnote{Email:  jutta.kunz@uni-oldenburg.de}
}
\affiliation{$^1$
Dept. Theor. and Nucl. Phys., KazNU, Almaty, 010008, Kazakhstan \\
$^2$ Institute for Basic Research,
Eurasian National University,
Astana, 010008, Kazakhstan
$^3$Institute of Physicotechnical Problems and Material Science of the NAS
of the
Kyrgyz Republic, 265 a, Chui Street, Bishkek, 720071,  Kyrgyz Republic \\
$^4$Institut f\"ur Physik, Universit\"at Oldenburg, Postfach 2503
D-26111 Oldenburg, Germany
}

\begin{abstract}
Cosmology with non-perturbative quantum corrections resulting from torsion is considered. It is shown that the evolution of closed, open and flat Universes is changed because of the presence of a non-zero dispersion of quantum torsion. The evolution of a Universe with  quantum torsion and with one type of average curvature can be similar to the evolution of a Universe without quantum torsion and with another type of average curvature. For the description of the non-perturbative quantum torsion, a vector field approximation is applied.
\end{abstract}

\maketitle

\section{Introduction}

The inclusion of torsion is a natural way for the generalization of the connection compatible with the metric. It leads to an affine connection constructed from Christoffel symbols plus a contortion tensor. Einstein-Cartan gravity (ECg) naturally extends general relativity to include the torsion. ECg is a part of non-Riemannian gravities. Nearly all
non-Riemannian models may be classified within the language of metric-affine
gravity (for review, see \cite{Hehl:1976kj}). 

ECg has some distinctive features as compared to general relativity. One can show that spin and torsion can avert cosmological singularities for certain spin configurations and for all configurations of matter with spin \cite{Kopczynski}. In Ref. \cite{Hehl1974} it was shown that those models violate an energy condition of a singularity theorem. In Ref. \citep{Odintsov} the account of quantum torsion to FRW cosmology was considered. In the textbook \cite{Blagojevic:2012bc} one can find exact cosmological and spherically symmetric solutions with torsion.

Modern experimental tests of the existence of torsion have led to negative results. In \cite{Kostelecky:2007kx,Heckel:2008hw,Shapiro:2001rz} tight bounds on many components of the torsion tensor (mixed-symmetry, trace and axial components) were reported, based on high-precision data from masers and torsion pendulums. In Ref. \cite{Shapiro:2001rz} various aspects of space-time torsion were considered, leading to the conclusion that there are no definite indications from experiments, whether torsion exists or not. On the other hand, theoretical studies give severe limits on torsion, since serious problems appear when one demands consistency of a propagating torsion theory at the quantum level. Therefore in our investigation we consider non-propagating and non-perturbative quantized torsion. Whereas the experiments gave negative results for the search of non-zero torsion, they had nothing to say for the case that torsion is a quantum field with zero expectation value but non-zero dispersion values.

In this letter we discuss the approximation when in the theory of gravity only the torsion is quantized, while the metric remains a classical object. This approximation will be used in the consideration of FRW cosmologies. Then such cosmologies represent non-Riemannian ones because torsion is included (for a review on non-Riemannian cosmologies see \cite{Puetzfeld:2004yg}.) After introducing some basic concepts of non-perturbative quantization (following Heisenberg), we use some approximation (vector field approximation) to obtain the Einstein equations with zero torsion and non-zero torsion dispersion. The vector field approximation means that we approximate the two point Green's function
$\left\langle
\hat Q_{\mu \nu \rho} (x)
\hat Q_{\alpha \beta \gamma} (y) \right\rangle $
by a product of a complex vector field $A_\mu$ in the spacetime points $x$ and $y$ (here $\hat Q$ denotes the quantum torsion operator).

\section{The basics of Heisenberg's non-perturbative quantization: the application to gravity}

According to Heisenberg \cite{heisenberg} the quantum operators of the metric
$\hat g_{\mu \nu}$ and the affine connection $\hat \Gamma^\rho_{\phantom{\rho} \mu \nu}$ obey the operator Einstein equations in the Palatini formalism
\begin{eqnarray}
	\hat \Gamma^\rho_{\phantom{\rho} \mu \nu} &=&
	\hat G^\rho_{\phantom{\rho} \mu \nu} +
	\hat K^\rho_{\phantom{\rho} \mu \nu} ,
\label{1-10}\\
	\hat R_{\mu \nu} - \frac{1}{2} \hat g_{\mu \nu} \hat R &=&
	\varkappa \hat T_{\mu \nu} ,
\label{1-20}\\
	\hat G^\rho_{\phantom{\rho} \mu \nu} &=&
	\frac{1}{2} \hat g^{\rho \sigma} \left(
		\frac{\partial \hat g_{\mu \sigma}}{\partial x^\nu} +
		\frac{\partial \hat g_{\nu \sigma}}{\partial x^\mu} -
		\frac{\partial \hat g_{\mu \nu}}{\partial x^\sigma}
	\right)
\label{1-30}
\end{eqnarray}
here $\varkappa=8 \pi G/c^4$; $\hat R_{\mu \nu}$ is the operator of the Ricci tensor; $\hat G^\rho_{\phantom{\rho} \mu \nu}$ are the operators of the Christoffel symbols; $\hat K^\rho_{\phantom{\rho} \mu \nu}$ is the operator of the contortion tensor,
$\hat T_{\mu \nu}$ is the operator of the matter fields and $\hat R$ is the operator of the scalar curvature defined in usual manner
\begin{eqnarray}
	\hat R_{\mu \nu} &=& \hat R^\rho_{\phantom{\rho} \mu \rho \nu},
\label{1-40}\\
	\hat R^\rho_{\phantom{\rho} \sigma \mu \nu} &=&
	\frac{\partial \hat \Gamma^\rho_{\phantom{\rho} \sigma \nu}}
	{\partial x^\mu} -
	\frac{\partial \hat \Gamma^\rho_{\phantom{\rho} \sigma \mu}}
	{\partial x^\nu} +
	\hat \Gamma^\rho_{\phantom{\rho} \tau \mu}
	\hat \Gamma^\tau_{\phantom{\tau} \sigma \nu} -
	\hat \Gamma^\rho_{\phantom{\rho} \tau \nu}
	\hat \Gamma^\tau_{\phantom{\tau} \sigma \mu} .
\label{1-50}
\end{eqnarray}
The operator of the contortion tensor $\hat K^\rho_{\phantom{\rho} \mu \nu}$ is defined via the operator of the torsion tensor
$\hat Q_{\mu \nu}^{\phantom{\mu \nu}\rho}$
\begin{equation}
	\hat K^\rho_{\phantom{\rho} \mu \nu} =
	\hat Q^\rho_{\phantom{\rho} \mu \nu} +
	\hat Q_{\mu \nu}^{\phantom{\mu \nu} \rho} -
	\hat Q_{\nu \phantom{\rho} \mu}^{\phantom{\mu} \rho}.
\label{1-60}
\end{equation}
The operator of the torsion tensor is the skew-symmetric part of the affine connection
\begin{equation}
	\hat Q_{\mu \nu}^{\phantom{\mu \nu} \rho} = \frac{1}{2} \left(
		\hat \Gamma_{\mu \nu}^{\phantom{\mu \nu} \rho} -
		\hat \Gamma_{\nu \mu}^{\phantom{\mu \nu} \rho}
	\right).
\label{1-70}
\end{equation}
Let us note that all quantities in Eqs.~\eqref{1-10} - \eqref{1-70} are operators. This leads to a difficult problem because now one has to solve the operator differential equations. Actually this is the main problem of nonperturbative quantization (that is well known.)

\textcolor{blue}{\textit{The non-perturbative quantization for Einstein gravity means that the quantum operators of metrics $\hat g_{\mu \nu}$, Christoffel symbols $\hat G^\rho_{\phantom{\rho} \mu \nu}$ and torsion $\hat Q_{\mu \nu}^{\phantom{\mu \nu} \rho}$ obey the operator Einstein - Cartan equations \eqref{1-10} - \eqref{1-60}.}}

\section{Vector field approximation for non-perturbative quantization of torsion}

In the approximation presented here we reserve the metric as a classical object and we quantize the torsion only.

For simplicity we will consider the skew-symmetric torsion
\begin{equation}
	\hat K_{\rho \mu \nu} = \hat Q_{\rho \mu \nu} = \hat Q_{[\rho \mu \nu]}.
\label{4-10}
\end{equation}
In our approach we consider the torsion with zero expectation value
\begin{equation}
	\left\langle
		\hat Q^\rho_{\phantom{\rho} \mu \nu}
	\right\rangle = 0
\label{4-20}
\end{equation}
but with non-zero dispersion
\begin{equation}
	\left\langle
		\left(
			\hat Q^\rho_{\phantom{\rho} \mu \nu}
		\right)^2
	\right\rangle \neq 0.
\label{4-30}
\end{equation}
We use a vector field approximation
\begin{equation}
	\left\langle
		\hat Q_{\rho_1 \mu_1 \nu_1}(x_1) \hat Q_{\rho_2 \mu_2 \nu_2}(x_2)
	\right\rangle \approx \epsilon_{\rho_1 \mu_1 \nu_1 \alpha}
	\epsilon_{\rho_2 \mu_2 \nu_2 \beta}
	A^\alpha (x_1) A^{\beta} (x_2).
\label{4-40}
\end{equation}
Now we have to make the following remark. The (classical and quantum) torsion can be split into three irreducible components:
\begin{equation}
	Q_{\alpha \beta \gamma} = \dfrac{1}{3} \left(
		Q_\beta g_{\alpha \gamma} - Q_\gamma g_{\alpha \beta}
	\right) - \dfrac{1}{6} \epsilon_{\alpha \beta\gamma \delta} S^\delta +
	q_{\alpha \beta \gamma}
\label{4-45}
\end{equation}
where $Q_\alpha = Q^\beta_{\cdot \alpha \beta}$ is the trace vector; the axial vector $S^\alpha = \epsilon^{\beta \gamma \delta \alpha} Q_{\beta \gamma \delta}$ and the tensor $q^\alpha_{\cdot \beta \gamma}$ satisfy the conditions $q^\alpha_{\cdot \beta \alpha = 0}$ and
$\epsilon^{\beta \gamma \delta \alpha} Q_{\beta \gamma \delta} = 0$. It is necessary to say that the vector $A_\mu$ has nothing to do with the vector $Q_\alpha$ or with the axial vector $S_\alpha$. Physically, $A_\mu$ represents some approximation for the expectation value \eqref{4-40}. This is not exact equality and it represents our physical assumption on the behaviour of the Green's function for torsion. In contrast the irreducible decomposition is an exact mathematical equality.

The expectation values of the Ricci and scalar curvature operators with the vector field approximation for the non-perturbative gravity quantization will be
\begin{eqnarray}
	\left\langle
		\hat R_{\mu \nu}
	\right\rangle &=& \tilde{R}_{\mu \nu} -
	\left\langle
		\hat Q^\rho_{\phantom{\rho} \mu \sigma}
		\hat Q^\sigma_{\phantom{\sigma} \rho \nu}
	\right\rangle = \tilde{R}_{\mu \nu} -
	\epsilon^\rho_{\phantom{\rho} \mu \sigma \alpha}
	\epsilon^\sigma_{\phantom{\sigma}\rho \nu \beta}
	A^\alpha A^\beta = \tilde{R}_{\mu \nu} + 2 \left(
		g_{\mu \nu} A_\alpha A^\alpha - A_\mu A_\nu
	\right) ,
\label{4-50}\\
	\left\langle	\hat R	\right\rangle &=&
	\tilde{R} + 6 A^\mu A_\mu,
\label{4-60}
\end{eqnarray}
where $\tilde{R}_{\mu \nu}$ and $\tilde{R}$ are the Ricci tensor and scalar curvature obtained from the metric in the standard way. Then the Einstein equations are
\begin{equation}
	\tilde{R}_{\mu \nu} -
		\frac{1}{2} g_{\mu \nu} \tilde{R} - \left(
		g_{\mu \nu} A_\alpha A^\alpha + 2 A_\mu A_\nu
	\right) = \varkappa T_{\mu \nu},
\label{4-70}
\end{equation}
where for simplicity we consider classical fields on the RHS. Now we would like to obtain an equation for the vector field $A_\mu$, approximately describing non-perturbative quantum gravitational effects. In order that the Einstein equations
\begin{equation}
	\left\langle
		\hat{R}_{\mu \nu} - \frac{1}{2} g_{\mu \nu} \hat R
	\right\rangle = \varkappa T_{\mu \nu}
\label{3b-105}
\end{equation}
are not overdetermined we demand that
\begin{equation}
	\left\langle
	\left(
		\hat{R}^\mu_\nu - \frac{1}{2} \delta^\mu_\nu \hat R
	\right)
	\right\rangle_{; \mu} = 0
\label{3b-107}
\end{equation}
since
\begin{equation}
	{T^\mu_\nu}_{; \mu} = 0.
\label{3b-108}
\end{equation}
Taking into account that
\begin{equation}
	\left( \tilde{R}^\mu_\nu - \frac{1}{2} \delta^\mu_\nu
	\tilde R \right)_{;\mu}	= 0
\label{3b-110}
\end{equation}
we obtain the desired equation for the vector field $A_\mu$
\begin{equation}
	\left(
		\delta^\mu_\nu A^\alpha A_\alpha + 2 A^\mu A_\nu
	\right)_{; \mu} = 0
\label{4-80}
\end{equation}
or
\begin{equation}
	\left( A^\alpha A_\alpha \right)_{, \nu} + 2 A^\mu_{; \mu} A_\nu +
	2 A^\mu A_{\nu ; \mu} = 0.
\label{4-90}
\end{equation}

\section{Non-perturbative quantum torsion in cosmology}

Now we would like to consider cosmology with quantum corrections coming from the torsion.We take the metric in the form
\begin{equation}
	ds^2 =
	a^2(\eta) \left\lbrace
   d\chi^2 - \left[ d \chi^2 +
   \left( \frac{\sin\sqrt{k}\chi}{\sqrt{k}} \right)^2
   \left(
    d\theta^2 + \sin^2 \theta d\varphi^2
   \right)
   \right]
  \right\rbrace
\label{eqs-10}
\end{equation}
with $k=+1, -1, 0$ for closed, open, and flat Universes, respectively.

We take the vector field $A_\mu$ as follows
\begin{equation}
	A_\mu = \left ( \phi (\eta), 0, 0, 0 \right ).
\label{eqs-20}
\end{equation}
The averaged Einstein equations \eqref{4-70} in the presence of matter in the form of dust are
\begin{eqnarray}
	\frac{{a'}^2}{a^2} + \left(
		k - \phi_0^2	
	\right) 	&=& \varkappa \frac{\varepsilon_0}{3 a},
\label{eqs-30}\\
	2 \frac{a''}{a} - \frac{{a'}^2}{a^2} + \left(
		k - \phi_0^2	
	\right) &=& 0,
\label{eqs-40}
\end{eqnarray}
where the prime denotes differentiation with respect to $\eta$; $\varepsilon_0$ is a constant. In Eqs.~\eqref{eqs-30}, \eqref{eqs-40} it is taken into account that Eq.~\eqref{4-90} with the ansatz \eqref{eqs-20} yields
\begin{equation}
	\phi' = 0 ,
\label{eqs-60}
\end{equation}
and the solution of this equation is
\begin{equation}
	\phi = \phi_0 = \text{const}.
\label{eqs-70}
\end{equation}
We see that the quantum torsion correction $\phi_0^2$ changes the evolution of the Universe by adding the term $-\phi_0^2$ to the Friedman equations. In hydrodynamical language, this term corresponds to a perfect fluid in the form of the so-called ``string gas'' or ``gas of strings'' having an equation of state
$p=-\varepsilon/3$.  Since in form the term with  $\phi_0^2$ is similar to that with $k$  (which describes the spatial curvature for the metric \eqref{eqs-10}), it is natural to introduce some geometrical interpretation
for this term \cite{Kamenshchik:2011jq}. Below we consider the consequences of the presence of this term on the evolution of three models corresponding to closed, open and flat Universes.


\subsection{Closed Universe}

The most interesting case is a closed Universe, $k=1$. Let us here consider three cases: the first case with $\phi_0^2 = 1$; the second one with $\phi_0^2 > 1$ and the third one with $\phi_0^2 < 1$:
\begin{itemize}
\item $\phi_0^2 = 1$. In this case Eq.~\eqref{eqs-30} tells us that the evolution of the scale factor $a(t)$ is the same as for a flat (non-torsion) Universe. The parametric dependence $a(t)$ is as follows
\begin{eqnarray}
	a(\eta) &=& \frac{\varkappa \varepsilon_0}{12} \eta^2 ,
\label{closed-13}\\
	ct(\eta) &=& \frac{\varkappa \varepsilon_0}{36} \eta^3 .
\label{closed-16}
\end{eqnarray}
The scale factor evolves as
\begin{equation}
	a = \left(
	 \frac{3 \varkappa \varepsilon_0 }{4}
	\right)^{1/3} (c t)^{2/3}.
\label{closed-10}
\end{equation}
Thus the evolution of a closed quantum-torsion Universe will not differ from the evolution of a flat (non-torsion) Universe.
\item $\phi_0^2 > 1$. In this case the term $1 - \phi_0^2$ in Eq.~\eqref{eqs-30} is negative and the closed Universe evolves as an open (non-torsion) Universe. The parametric dependence $a(t)$ is as follows
\begin{eqnarray}
	a(x) &=& \frac{\varkappa \varepsilon_0}{6 (\phi_0^2 - 1)}
	\left(
		\cosh x  -1
	\right),
\label{closed-20}\\
	t(x) &=& \frac{\varkappa \varepsilon_0}{6 (\phi_0^2 - 1) c}
	\left (
		\sinh x  - x
	\right),
\label{closed-30}
\end{eqnarray}
where $x = \eta \sqrt{\phi_0^2 - 1}$. Considering large values of
$x \gg 1$ yields
\begin{equation}
	a \approx c t .
\label{closed-35}
\end{equation}
This expression coincides with the expression for an open (non-torsion) Universe.
In turn, close to the Big Bang, i.e., when $x\ll 1$,
\begin{equation}
a \approx 3 \left[\frac{\varkappa \varepsilon_0}{36(\phi_0^2 - 1)}\right]^{1/3}(c t)^{2/3},
\label{closed-36}
\end{equation}
i.e., one has the expression similar to that of the case of open Universe
 without the quantum torsion.
\item $\phi_0^2 < 1$. In this case the term $1 - \phi_0^2$ in equation \eqref{eqs-30} is positive and the closed Universe evolves similar to the closed (non-torsion) Universe. Here the parametric dependence $a(t)$ is the following
\begin{eqnarray}
	a(x) &=& \frac{\varkappa \varepsilon_0}{6 (1 - \phi_0^2)}
	\left(
		1 - \cos x
	\right),
\label{closed-40}\\
	t(x) &=& \frac{\varkappa \varepsilon_0}{6 (1 - \phi_0^2) c}
	\left(
		x - 	\sin x
	\right),
\label{closed-50}
\end{eqnarray}
where $x = \eta \sqrt{1 - \phi_0^2}$. Again, close to the Big Bang $a(t)$ is
\begin{equation}
a \approx 3 \left[\frac{\varkappa \varepsilon_0}{36(1-\phi_0^2)}\right]^{1/3}(c t)^{2/3}.
\label{closed-51}
\end{equation}
\end{itemize}

\subsection{Open Universe}

In this case the parameter $k=-1$ and the quantum torsion correction $- \phi_0^2$ are both negative. In this case the parametric dependence $a(t)$ is the following
\begin{eqnarray}
	a(x) &=& \frac{\varkappa \varepsilon_0}{6 (\phi_0^2 + 1)}
	\left(
		\cosh x  -1
	\right),
\label{open-10}\\
	t(x) &=& \frac{\varkappa \varepsilon_0}{6 (\phi_0^2 + 1)c}
	\left(
		\sinh x  - x
	\right),
\label{open-20}
\end{eqnarray}
where $x = \eta \sqrt{\phi_0^2 + 1}$. For large values of
$x \gg 1$
\begin{equation}
	a \approx c t .
\label{open-30}
\end{equation}
For $x<<1$ we have
\begin{equation}
a \approx 3 \left[\frac{\varkappa \varepsilon_0}{36(\phi_0^2 + 1)}\right]^{1/3}(c t)^{2/3}.
\label{open-31}
\end{equation}

\subsection{Flat Universe}

In this case the parameter $k = 0.$ The parametric dependence $a(t)$ is the following
\begin{eqnarray}
	a(x) &=& \frac{\varkappa \varepsilon_0}{6 \phi_0^2 }
	\left[
		\cosh x  - 1
	\right],
\label{flat-10}\\
	t(x) &=& \frac{\varkappa \varepsilon_0}{6 \phi_0^2 c}
	\left [
		\sinh x  - x
	\right],
\label{flat-20}
\end{eqnarray}
where $x = \eta \phi_0$. For large values of
$x \gg 1$
\begin{equation}
	a \approx c t .
\label{flat-30}
\end{equation}
Close to the Big Bang (where $x \ll 1$) $a(t)$ is
\begin{equation}
a \approx 3 \left[\frac{\varkappa \varepsilon_0}{36\phi_0^2}\right]^{1/3}(c t)^{2/3}.
\label{flat-40}
\end{equation}

\section{Conclusions}

In this paper we have investigated the role of quantum torsion on the evolution of the Universe. We have considered a form of torsion with zero expectation value but with non-zero dispersion. We have used the approximation where the expectation value of the product of the torsion operator in two points is \textit{approximated} by a vector field.  As a result, we have shown that the quantum torsion may lead to a qualitative change of the evolution of the Universe. For example, a closed Universe with fluctuating quantum torsion may have an evolution similar to a closed, open or flat (non-torsion) Universe, depending on the value of the quantum fluctuation dispersion of the torsion. One can note that using a ``scalar approximation'' for the quantum torsion leads to the dark energy \cite{Dzhunushaliev:2012nf}.

\section*{Acknowledgements}
V.D. and V.F. are grateful to the Research Group Linkage Programme of the Alexander von Humboldt Foundation for the support of this research. They also would like to thank the Carl von Ossietzky University of Oldenburg for hospitality while this work was carried out. B.K. and J.K. gratefully acknowledge support by the DFG Research Training Group 1620 ``Models of Gravity''. This work is partially supported by a grant in fundamental research in natural sciences by the Ministry of Education and Science of Kazakhstan.

\end{document}